\documentclass[aps,prl,twocolumn,superscriptaddress]{revtex4}
\usepackage{graphicx}

\begin{document}

\title{Coherent Quantum Dynamics of a Superconducting Flux Qubit}

\author{I. Chiorescu}
\altaffiliation{e-mail: chiorescu@qt.tn.tudelft.nl}
\affiliation{Quantum Transport Group, Department of NanoScience,
Delft University of Technology and Delft Institute for Micro
Electronics and Submicron Technology (DIMES), Lorentzweg 1, 2628
CJ Delft, Netherlands}
\author{Y. Nakamura}
\affiliation{Quantum Transport Group, Department of NanoScience,
Delft University of Technology and Delft Institute for Micro
Electronics and Submicron Technology (DIMES), Lorentzweg 1, 2628
CJ Delft, Netherlands}
\affiliation{NEC Fundamental Research
Laboratories, 34 Miyukigaoka, Tsukuba, Ibaraki 305-8501, Japan}
\author{C. J. P. M. Harmans}
\affiliation{Quantum Transport Group, Department of NanoScience,
Delft University of Technology and Delft Institute for Micro
Electronics and Submicron Technology (DIMES), Lorentzweg 1, 2628
CJ Delft, Netherlands}
\author{J. E. Mooij}
\affiliation{Quantum Transport Group, Department of NanoScience,
Delft University of Technology and Delft Institute for Micro
Electronics and Submicron Technology (DIMES), Lorentzweg 1, 2628
CJ Delft, Netherlands}

\date{submitted 2 December 2002; accepted 4 February 2003: \emph{Science} \textbf{299}, 1869 (2003)}

\begin{abstract}
We have observed coherent time evolution between two quantum
states of a superconducting flux qubit comprising three Josephson
junctions in a loop. The superposition of the two states carrying
opposite macroscopic persistent currents is manipulated by
resonant microwave pulses. Readout by means of switching-event
measurement with an attached superconducting quantum interference
device revealed quantum-state oscillations with high fidelity.
Under strong microwave driving it was possible to induce hundreds
of coherent oscillations. Pulsed operations on this first sample
yielded a relaxation time of 900 nanoseconds and a free-induction
dephasing time of 20 nanoseconds. These results are promising for
future solid-state quantum computing.
\end{abstract}

\maketitle

It is becoming clear that artificially fabricated solid-state
devices of macroscopic size may, under certain conditions, behave
as single quantum particles. We report on the controlled
time-dependent quantum dynamics between two states of a
micron-size superconducting ring containing billions of Cooper
pairs \emph{(1)}. From a ground state in which all the Cooper
pairs circulate in one direction, application of resonant
microwave pulses can excite the system to a state where all pairs
move oppositely, and make it oscillate coherently between these
two states. Moreover, multiple pulses can be used to create
quantum operation sequences. This is of strong fundamental
interest because it allows experimental studies on decoherence
mechanisms of the quantum behavior of a macroscopic-sized object.
In addition, it is of great significance in the context of quantum
computing \emph{(2)} because these fabricated structures are
attractive for a design that can be scaled up to large numbers of
quantum bits or qubits \emph{(3)}.

Superconducting circuits with mesoscopic Josephson junctions are
expected to behave according to the laws of quantum mechanics if
they are separated sufficiently from external degrees of freedom,
thereby reducing the decoherence. Quantum oscillations of a
superconducting two-level system have been observed in the Cooper
pair box qubit using the charge degree of freedom \emph{(4)}. An
improved version of the Cooper pair box qubit showed that quantum
oscillations with a high quality factor could be achieved
\emph{(5)}. In addition, a qubit based on the phase degree of
freedom in a Josephson junction was presented, consisting of a
single, relatively large Josephson junction current-biased close
to its critical current \emph{(6,7)}.

Our flux qubit consists of three Josephson junctions arranged in a
superconducting loop threaded by an externally applied magnetic
flux near half a superconducting flux quantum $\Phi_0=h/2e$
[\emph{(8)}; a one-junction flux-qubit is described in
\emph{(9)}]. Varying the flux bias controls the energy level
separation of this effectively two-level system. At half a flux
quantum, the two lowest states are symmetric and antisymmetric
superpositions of two classical states with clockwise and
anticlockwise circulating currents. As shown by previous microwave
spectroscopy studies, the qubit can be engineered such that the
two lowest eigenstates are energetically well separated from the
higher ones \emph{(10)}. Because the qubit is primarily biased by
magnetic flux, it is relatively insensitive to the charge noise
that is abundantly present in circuits of this kind.

The central part of the circuit, fabricated by electron beam
lithography and shadow evaporation of Al, shows the three in-line
Josephson junctions together with the small loop defining the
qubit in which the persistent current can flow in two directions,
as shown by arrows (Fig.~1A). The area of the middle junction of
the qubit is $\alpha=0.8$ times the area of the two outer ones.
This ratio, together with the charging energy $E_C=e^2/2C$ and the
Josephson energy $E_J=hI_C/4\pi e$ of the outer junctions (where
$I_C$ and $C$ are their critical current and capacitance,
respectively), determines the qubit energy levels (Fig. 2A) as a
function of the superconductor phase $\gamma_q$ across the
junctions (Fig.~1B). Close to $\gamma_q=\pi$, the loop behaves as
a two-level system with an energy separation $E_{10}=E_1-E_0$ of
the eigenstates $|0\rangle$ and $|1\rangle$ described by the
effective Hamiltonian $H=-\epsilon\sigma_z/2-\Delta\sigma_x/2$,
where $\sigma_{z,x}$ are the Pauli spin matrices, $\Delta$ is the
level repulsion, and $\epsilon\approx I_p\Phi_0(\gamma_q-\pi)/\pi$
(where $I_p\approx 2\pi\alpha E_J/\Phi_0$ is the qubit maximum
persistent current) \emph{(11)}.

\begin{figure}
\includegraphics[width=8.5cm]{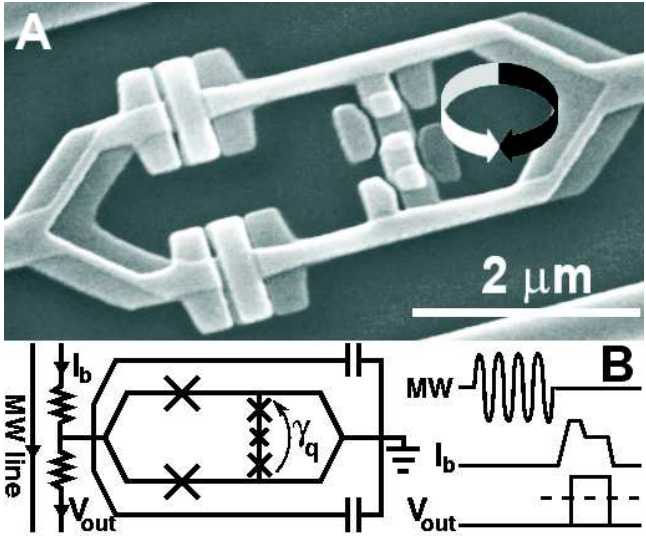}
\caption{\textbf{(A)} Scanning electron micrograph of a flux-qubit
(small loop with three Josephson junctions of critical current
$\sim$0.5~ mA) and the attached SQUID (large loop with two big
Josephson junctions of critical current $\sim$2.2~mA). Evaporating
Al from two different angles with an oxidation process between
them gives the small overlapping regions (the Josephson
junctions). The middle junction of the qubit is 0.8 times the area
of the other two, and the ratio of qubit/SQUID areas is about 1:3.
Arrows indicate the two directions of the persistent current in
the qubit. The mutual qubit/SQUID inductance is $M\approx9$~pH.
\textbf{(B)} Schematic of the on-chip circuit; crosses represent
the Josephson junctions. The SQUID is shunted by two capacitors
($\sim$5~pF each) to reduce the SQUID plasma frequency and biased
through a resistor ($\sim$150~ohms) to avoid parasitic resonances
in the leads. Symmetry of the circuit is introduced to suppress
excitation of the SQUID from the qubit-control pulses. The MW line
provides microwave current bursts inducing oscillating magnetic
fields in the qubit loop. The current line provides the measuring
pulse $I_b$ and the voltage line allows the readout of the
switching pulse $V_{out}$. The $V_{out}$ signal is amplified, and
a threshold discriminator (dashed line) detects the switching
event at room temperature.}
\end{figure}

The sample is enclosed in a gold-plated copper shielding box kept
at cryogenic temperatures T = 25~mK (k$_B$T $\ll\Delta$). The
qubit is initialized to the ground state simply by allowing it to
relax. Coherent control of the qubit state is achieved by applying
resonant microwave excitations on the microwave (MW) line
(Fig.~1B), thereby inducing an oscillating magnetic field through
the qubit loop. The qubit state evolves driven by a time-dependent
term $(-1/2)\epsilon_{mw}\cos(2\pi Ft)\sigma_z$ in the Hamiltonian
where \emph{F} is the microwave frequency and $\epsilon_{mw}$ is
the energy-modulation amplitude proportional to the microwave
amplitude. This dynamic evolution is similar to that of spins in
magnetic resonance. When the MW frequency equals the energy
difference of the qubit, the qubit oscillates between the ground
state and the excited state. This phenomenon is known as Rabi
oscillation. The Rabi frequency depends linearly on the MW
amplitude \emph{(12-14)}.

Readout is performed with an underdamped superconducting quantum
interference device (SQUID) with a hysteretic current-voltage
characteristic in direct contact with the qubit loop (Fig.~1A).
The mutual coupling $M$ is relatively large because of the shared
kinetic and geometric inductances of the joint part enhancing the
qubit signal. After performing the qubit operation, a bias current
pulse $I_b$ is applied to the SQUID \emph{(15)}. The $I_b$ pulse
consists of a short current pulse of length $\sim$50~ns followed
by a trailing plateau of $\sim$500~ns (Fig.~1B). During the
current pulse, the SQUID either switches to the gap voltage or
stays at zero voltage. The pulse height and length are set to
optimize the distinction of the switching probability between the
two qubit states, which couple to the SQUID through the associated
circulating currents. Because the readout electronics has a
limited bandwidth of $\sim$100~kHz, a voltage pulse of 50~ns pulse
is too short to be detected. For that reason the trailing plateau
is added, with a current just above the retrapping current of the
SQUID. The whole shape is adjusted for maximum readout fidelity.
The switching probability is obtained by repeating the whole
sequence of reequilibration, microwave control pulses and readout
typically 5000 times.

\begin{figure}
\includegraphics[width=8.5cm]{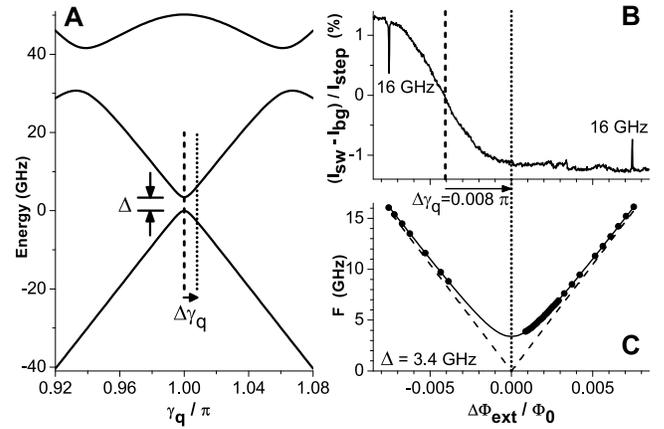}
\caption{\textbf{(A)} Calculated energy diagram for the
three-junction qubit, for $E_J/E_C=35$, $E_C=7.4$~GHz and $\alpha=
0.8$ \emph{(11)}. $\Delta\gamma_q$ indicates the phase shift
induced by the SQUID bias current. \textbf{(B)} Ground-state
transition step: The sinusoidal background modulation of the SQUID
($I_{bg}$) is subtracted from the $I_b$ pulse amplitude
corresponding to 50$\%$ switching probability ($I_{sw}$) and then
normalized to $I_{step}$, the middle value (at the dashed line). A
sharp peak and dip are induced by a long (1 $\mu$s) MW radiation
burst at 16 GHz, allowing the symmetry point to be found (midpoint
of the peak/dip positions, dotted line). Data show $I_{sw}$ versus
$\Delta\Phi_{ext}$, the deviation in external flux from this
point. The transition step is displaced from this point by
$\Delta\gamma_q/2\pi$. \textbf{(C)} Frequency of the resonant
peaks/dips (dots) versus $\Delta\Phi_{ext}$; the continuous line
is a numerical fit with the same parameters as in \textbf{(A)}
leading to a value of $\Delta=3.4$~GHz, whereas the dashed line
depicts the case $\Delta=0$.}
\end{figure}

When the SQUID bias current is switched on, the circulating
current in the SQUID changes. This circulating current, coupled to
the qubit through the mutual inductance, changes the phase bias of
the qubit by an estimated amount $0.01\pi$. Consequently the phase
bias at which the quantum operations are performed is different
from the phase bias at readout. This can be very useful because at
the phase bias near $\pi$, where the qubit is least sensitive to
flux noise, the expectation values for the qubit circulating
current are extremely small. The automatic phase bias shift can be
used to operate near $\pi$ and to perform readout at a bias with a
good qubit signal \emph{(11)}. Care must be taken that the fast
shift remains adiabatic and that the whole sequence is completed
within the relaxation time.

\begin{figure}
\includegraphics[width=8.5cm]{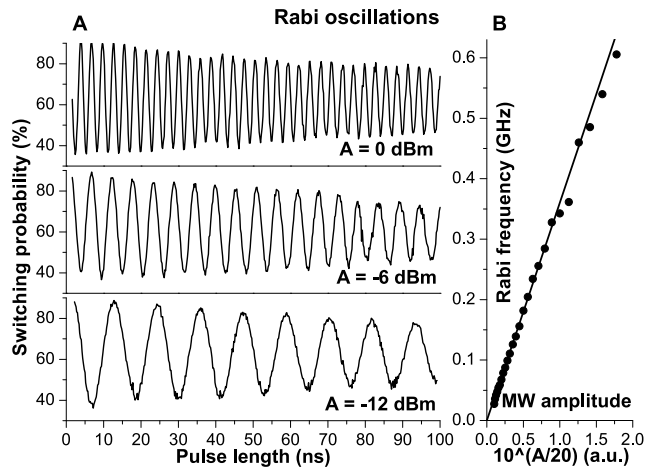}
\caption{\textbf{(A)} Rabi oscillations for a resonant frequency
$F=E_{10}=6.6$~GHz and three different microwave powers $A=0$, -6
and -12 dBm, where $A$ is the nominal microwave power applied at
room temperature. The data are well fitted by exponentially damped
sinusoidal oscillations. The resulting decay time is $\sim$150~ns
for all powers. \textbf{(B)} Linear dependence of the Rabi
frequency on the microwave amplitude, expressed as 10$^{A/20}$.
The slope is in agreement with estimations based on sample
design.}
\end{figure}

The average SQUID switching current $I_{sw}$ versus applied flux
shows the change of the qubit ground-state circulating current
(Fig.~2B). Here, the $I_b$ pulse amplitude is adjusted such that
the averaged switching probability is maintained at 50$\%$. A step
corresponding to the change of qubit circulating current was
observed (around the dashed line). The relative variation of
2.5$\%$ of $I_{sw}$ is in agreement with the estimation based on
the qubit current $I_p$ and the qubit-SQUID mutual inductance $M$.

The relevant two energy levels of the qubit were first examined by
spectroscopic means. Before each readout, a long microwave pulse
(1~$\mu$s) at a series of frequencies was applied to observe
resonant absorption peaks/dips each time the qubit energy
separation $E_{10}$ $-$ adjusted by changing the external flux $-$
coincides with the MW frequency $F$ \emph{(10)}. The dots in
Fig.~2C are measured peak/dip positions, obtained by varying $F$,
whereas the continuous line is a numerical fit produced by exact
diagonalization (compare Fig.~2A) giving an energy gap
$\Delta\approx3.4$~GHz. The curves in Fig.~2, B and C, are plotted
against the change $\Delta\Phi_{ext}$ in external flux from the
symmetry position indicated by the dotted line. In agreement with
our numerical simulations, the step (Fig.~2B) is shifted away from
the symmetry position of the energy spectrum (Fig.~2C) by a phase
bias shift
$\Delta\gamma_q\approx2\pi(\Delta\Phi_{ext}/\Phi_0)\approx0.008\pi$.
The step reflects the external-flux dependence of the qubit
circulating current at $I_b\approx I_{sw}$ (after the shift),
whereas the spectrum reflects $E_{01}$ at $I_b = 0$ (before the
shift) \emph{(16)}.

\begin{figure}
\includegraphics[width=8.5cm]{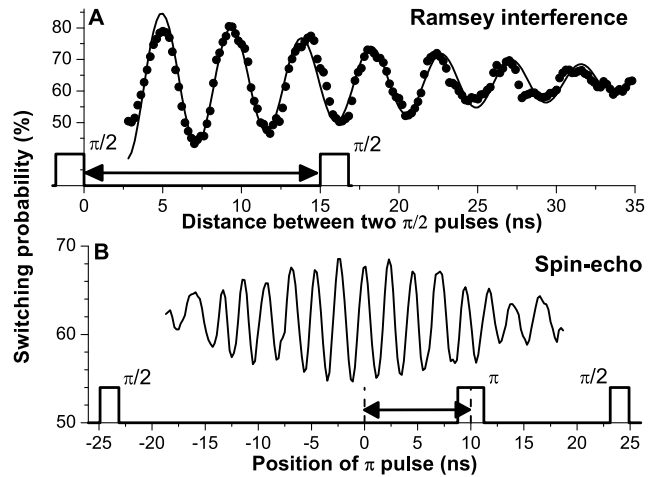}
\caption{\textbf{(A)} Ramsey interference: The measured switching
probability (dots) is plotted against the time between the two
$\pi/2$ pulses. The continuous line is a fit by exponentially
damped oscillations with a decay time of 20~ns. The Ramsey
interference period of 4.5~ns agrees with the inverse of the
detuning from resonance, 220~MHz. The resonant frequency is
5.71~GHz and microwave power $A=0$~dBm. \textbf{(B)} Spin-echo
experiment: switching probability versus position of the $\pi$
pulse between two $\pi/2$ pulses. The period of $\sim$2.3~ns
corresponds well to half the inverse of the detuning. The width
and timing of microwave pulses in the MW line are shown in each
graph. The readout pulse in the bias line immediately follows the
last $\pi/2$ pulse (see Fig.~1B).}
\end{figure}

Next, we used different MW pulse sequences to induce coherent
quantum dynamics of the qubit in the time domain. For a given
level separation $E_{10}$, a short resonant MW pulse of variable
length with frequency $F = E_{10}$ was applied. Together with the
MW amplitude, the pulse length defines the relative occupancy of
the ground state and the excited state. Corresponding switching
probability was measured with a fixed-bias current pulse
amplitude. We obtained coherent Rabi oscillations of the qubit
circulating current for a frequency $F = 6.6$~GHz and three
different values of the MW power $A$ (Fig.~3). The variation in
switching probability is around 60$\%$, indicating that the
fidelity in a single readout is of that order. By varying $A$, we
verified the linear dependence of the Rabi frequency on the MW
amplitude, a key signature of the Rabi process (Fig.~3B). The
oscillation pattern can be fitted to a damped sinusoid. For
relatively strong driving (Rabi period below 10~ns), decay times
$\tau_{Rabi}$ up to $\sim$150~ns are obtained. This large decay
time resulted in hundreds of coherent oscillations at large
microwave power.

The Rabi scheme also allows the study of the state occupancy
relaxation. This can be done by applying a coherent $\pi$ pulse
for full rotation of the qubit into the excited state and varying
the delay time before readout. Experiments performed at
$F=5.71$~GHz gave an exponential decay with relaxation time
$\tau_{relax}\approx900$~ns.

As a next step we measured the undriven, free-evolution dephasing
time $\tau_\phi$ by performing a Ramsey interference experiment
\emph{(17)} as follows. Two $\pi/2$ pulses, whose length is
determined from the Rabi precession presented above, are applied
to the qubit. The first pulse creates a superposition of the
$|0\rangle$ and $|1\rangle$ states. If the microwave frequency is
detuned by $\delta F = E_{10}-F$ away from resonance, the
superposition phase increases with a rate $2\pi\delta F$, in the
frame rotating with the MW frequency $F$. After a varying delay
time, we apply a second $\pi/2$ pulse to measure the final
$|0\rangle$ and $|1\rangle$ state occupancy via the switching
probability. The readout shows Ramsey fringes with a period
$1/\delta F$, as in Fig.~4A, where $E_{10} = 5.71$~GHz and $\delta
F = 220$~MHz. The dots represent experimental data, whereas the
continuous line is an exponentially damped sinusoidal fitting
curve, yielding a free-evolution dephasing time
$\tau_\phi\approx20$~ns. Note that the oscillation period of
4.5~ns agrees well with $1/\delta F$.

Additional information on the spectral properties of the
decohering fluctuations can be obtained with a modified Ramsey
experiment. By inserting a $\pi$ pulse between the two $\pi/2$
pulses (Fig.~4B), we obtain a spin-echo pulse configuration. The
role of the $\pi$ pulse is to reverse the noise-driven diffusion
of the qubit phase at the midpoint in time of the free evolution.
Dephasing due to fluctuations of lower frequencies should be
cancelled by their opposite influence before and after the $\pi$
pulse \emph{(18)}. Spin-echo oscillations (Fig.~4B) are taken
under the same conditions as the Ramsey fringes, but are here
recorded as a function of the $\pi$ pulse position. The period
($\sim$2.3~ns) is half that of the Ramsey interference. We
measured the decay of the maximum spin-echo signal (\emph{i.e.}
with the $\pi$ pulse in the center) versus the delay time between
the two $\pi/2$ pulses. The data can be fitted to a half-Gaussian
(not shown) with a decay time $\tau_{echo}\approx30$~ns.

We conclude that with the present device and setup, the dephasing
time $\tau_\phi\approx20$~ns, as measured with the Ramsey pulses,
is much shorter than the relaxation time
$\tau_{relax}\approx900$~ns. Dephasing is probably caused by a
variation in time of the qubit energy splitting, due to external
or internal noise. A likely source is external flux noise, which
can be reduced in the future. The present qubit could not be
operated at the symmetry point $\gamma_q=\pi$ where the influence
of flux noise is minimal \emph{(5)}, presumably as the result of
an accidentally close SQUID resonance \emph{(19)}. Other possible
noise sources are thermal, charge, critical current and spin
fluctuations. From estimations of the Johnson noise in the bias
circuit \emph{(20,21)}, we find a contribution that is several
orders of magnitude weaker.

For strong driving, Rabi oscillations persisted for times much
longer than $\tau_\phi$. This constitutes no inconsistency. The
dependence of the Rabi period on the detuning, due to fluctuations
of the qubit energy $E_{10}$, is weak when the Rabi period is
short. The fact that coherence is only marginally improved by the
$\pi$ pulse in the spin-echo experiment seems to indicate the
presence of noise at frequencies beyond 10~MHz. Further analysis
and additional measurements are needed.

These first results on the coherent time evolution of a flux qubit
are very promising. The already high fidelity of qubit excitation
and readout can no doubt be improved. Quite likely it is also
possible to reduce the dephasing rate. Taken together, these
results establish the superconducting flux qubit as an attractive
candidate for solid-state quantum computing.


\begin{thebibliography}{99}
\bibitem{1} A. J. Leggett, A. Garg, \textit{Phys. Rev. Lett.} \textbf{54}, 857
(1985).

\bibitem{2} M. A. Nielsen, I. L. Chuang, \textit{Quantum Computation and
Quantum Information} (Cambridge Univ. Press, Cambridge, 2000).

\bibitem{3} Y. Makhlin, G. Schön, A. Shnirman, \textit{Rev. Mod. Phys.} \textbf{73},
357 (2001).

\bibitem{4} Y. Nakamura, Yu. A. Pashkin, J.S. Tsai, \textit{Nature} \textbf{398}, 786 (1999).

\bibitem{5} D. Vion \textit{et al.}, \textit{Science} \textbf{296}, 886 (2002).

\bibitem{6} Y. Yu, S. Han, X.
Chu, S.-I. Chu, Z. Wang, \textit{Science} \textbf{296}, 889
(2002).

\bibitem{7} J. M. Martinis, S. Nam, J. Aumentado, C. Urbina, \textit{Phys. Rev.
Lett.} \textbf{89}, 117901 (2002).

\bibitem{8} J. E. Mooij \textit{et al.}, \textit{Science} \textbf{285}, 1036 (1999).

\bibitem{9} J. R. Friedman, V. Patel, W. Chen, S.K. Tolpygo, J.E. Lukens,
\textit{Nature} \textbf{406}, 43 (2000).

\bibitem{10} C. H. van der Wal \textit{et al.}, \textit{Science} \textbf{290}, 773 (2000).

\bibitem{11} The two opposite persistent currents states, depicted by
arrows in Fig.~1A, describe the basis ${|\!\!\uparrow\rangle,
|\!\!\downarrow\rangle}$ of Pauli spin matrices. Using the
notation $\tan2\theta = \Delta/e$, the qubit eigenstates can be
written as $|0\rangle =
\cos\theta|\!\!\uparrow\rangle+\sin\theta|\!\!\downarrow\rangle$
and $|1\rangle =
-\sin\theta|\!\!\uparrow\rangle+\cos\theta|\!\!\downarrow\rangle$
and the expectation values of the corresponding circulating
currents as $I_{q0,1}=\pm I_p\cos2\theta$.

\bibitem{12} I. I. Rabi, \textit{Phys. Rev.} \textbf{51}, 652 (1937).

\bibitem{13} M. Grifoni, P. Hänggi, \textit{Phys. Rep.} \textbf{304}, 229 (1998).

\bibitem{14} In the frame rotating at the MW frequency $F=E_{10}$, at the symmetry point
 ($\gamma_q=\pi$), the Rabi precession is around an axis perpendicular to x axis with a frequency
$(\epsilon_{mw}\sin2\theta)/h$ [with $\theta$ as in
\textit{(11)}].

\bibitem{15} During the qubit initialization and control, the SQUID bias
current is set to zero and, as a result of the SQUID symmetry, the
qubit is decoupled from the external current noise to first order.
At $I_b=0$, small external noise current flows equally in the two
branches of the SQUID even in the presence of the circulating
current in the SQUID. See also \textit{(21)}.

\bibitem{16} A part of the energy spectrum is missing, because the readout
is not efficient around the step and thus the spectroscopy signal
is weak.

\bibitem{17} N. F. Ramsey, \textit{Phys. Rev.} \textbf{78}, 695 (1950).

\bibitem{18} E. L. Hahn, \textit{Phys. Rev.} \textbf{80}, 580 (1950).

\bibitem{19} In the present device, $\Delta\approx3.4$~GHz was rather close to the
SQUID plasma frequency designed to be $\sim$2 GHz (at $I_b\approx
I_{sw}$). This could be a possible explanation for the absence of
coherent oscillations for $F = \Delta$.

\bibitem{20} Lin Tian, S. Lloyd, T.P. Orlando, \textit{Phys. Rev. B} \textbf{65}, 144516
(2002).

\bibitem{21} C.H. van der Wal, F.K. Wilhelm, C.J.P.M. Harmans, J.E. Mooij,
\textit{Eur. Phys. J. B} \textbf{31}, 111 (2003).

\bibitem{22} We thank R. N. Schouten, J. B. Majer, A. Lupascu, and K.
Semba, for experimental help and discussion; D. Esteve and C.
Urbina for valuable input; and D. Vion, A. Aassime, C. H. van der
Wal, A. C. J. ter Haar, T. Orlando, S. Lloyd, M. Grifoni, F. K.
Wilhelm, L. Vandersypen, and P.C.E. Stamp for fruitful
discussions. Supported by the Dutch Foundation for Fundamental
Research on Matter (FOM), the European Union SQUBIT project, and
the U.S. Army Research Office (grant DAAD 19-00-1-0548).


\end{thebibliography}
\end{document}